\documentclass[%
 reprint,
 amsmath,amssymb,
 aps,
]{revtex4-1}
\usepackage{mathrsfs}
\usepackage{graphicx}
\usepackage{dcolumn}
\usepackage{bm}

\begin{document}
\title{Absorption interferometer based on phase modulation}

\author{Miaodi Guo}
\author{Xuemei Su}
 \email{suxm@jlu.edu.cn}

\affiliation{%
 $^{1}$Key Lab of Coherent Light, Atomic and Molecular Spectroscopy, Ministry of Education; and College of Physics, Jilin University, Changchun 130012, People's Republic of China}%

\date{}

\begin{abstract}
We propose a scheme in which an arbitrary incidence can be made perfectly reflected/transmitted if a phase setup is adjusted under a specific condition. We analyze the intracavity field variation as well as the output field with changing closed-loop phase $\phi_{1}$ of atomic system and relative phase $\phi_{2}$ of input probe beams. And we obtain the condition for perfect transmitter or reflector. By adjusting two phase setups, the medium absorption and light interference can be controlled so that photon escape from cavity can be modulated, thus the intensity switching based on phase control can be realized. Then based on the transmission/reflection analysis, total absorption of this system can be investigated. Therefore our scheme can be used as an absorption interferometer to explore the optical absorption in some complicated system. The phase delay dependent on $\phi_{1}$ or $\phi_{2}$ in output light intensity can be applied in the realization of quantum phase gate and subtle wave filter. And based on this scheme, we implement the state transfer between perfect transmitter/reflector and non-perfect coherent photon absorber via relative-phase modulation.
\newline
~~~~~\newline
PACS number(s):42.50.-p, 42.50.Pq, 32.80.Qk, 42.25.Bs
\end{abstract}

\maketitle

\section{\normalsize\uppercase{Introduction}}
Optical interference is a direct evidence for wave theory of light, based on which many precise instruments and technologies have been applied, e.g. Michelson interferometer, Mach-Zehnder interferometer, interferometry and holography. With the development of quantum mechanics and cavity quantum electrodynamics (cavity-QED), quantum interference, which is based on the uncertainty of different transition channels, is becoming more and more important in the interactions of light and matter. One famous physical phenomenon is electromagnetically induced transparency (EIT), which can be controlled by manipulating the intensity or frequency of incident laser or density of matter \cite{PRA51/576,OL23/295,PRL100/173602,Nature465/755}. Therefore, many researches about light switching based on EIT have become hot topics \cite{PRL81/3611,PRA82/033808,PRA84/043821,PRA86/033828,PRA87/053802,PRA89/023806}. When EIT system is with a closed loop, the phenomenon becomes more interesting, where EIT shows phase-dependent property \cite{PRA59/2302,PLA324/388,PRA71/011803,MOP54/2459,CPB26/074207}. That means EIT effect can be manipulated in a more sophisticated way. Besides light switch \cite{PRA73/011802}, there are some new explorations based on phase-dependent EIT, such as phase-control spontaneous emission in EIT medium \cite{PRL81/293}, beam splitter \cite{PRL101/043601} and entanglement between collective fields \cite{PRA81/033836}. By controlling the phase shift of input lasers in atomic system, one can operate the nonadiabatic optical transitions and quantum mechanical superposition states \cite{PRA79/025401,PRA87/013430}. With the aid of Kerr cross-phase modulation, one can also realize the polarization selection in EIT medium \cite{PRA90/063841,PRA92/043838}. Phase shift also plays a vital role in typical optical interference. When put wave interference together with quantum system, the manipulation of optical absorption can reach to a new era \cite{PRL105/053901,Science331/889,PRA92/023824} and the system can be operated as a transmitter (reflector) or an absorber \cite{PRA95/013841}.

Here we propose a scheme to manipulate optical intensity in cavity-QED system via phase control. By using the closed loop and two coherent incident beams, we can manipulate the field intensity through the interaction of medium absorption and wave interference. The intracavity field and output field can be periodically modulated by relative phase and two output channels can be operated at the same pace or at fixed phase delay about relative phase. We make the theoretical analysis of this cavity-QED system and obtain the solutions for intracavity and output light field, respectively. We also explain the principle for our scheme to be acted as an absorption interferometer and an optical switching based on phase modulation in Sec. \uppercase\expandafter{\romannumeral2}. The detailed theoretical simulated results are presented in Sec. \uppercase\expandafter{\romannumeral3}. At last, we make a simple summary in Sec. \uppercase\expandafter{\romannumeral4}.

\section{\normalsize\uppercase{theoretical analysis}}
\begin{figure}[!hbt]
\centering
\includegraphics[width=6cm]{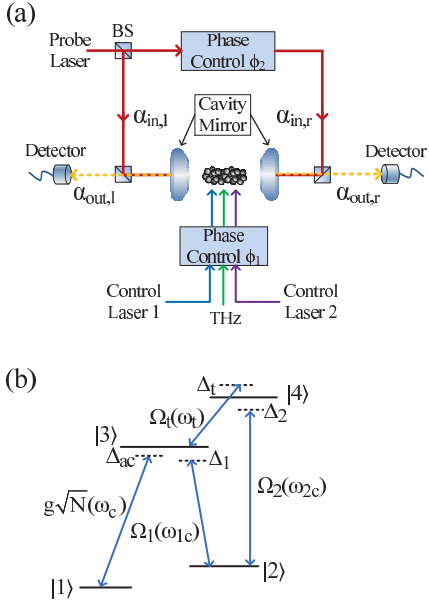}
\caption{(a) The theoretical scheme and (b) the four-level atomic level structure of $^{87}$Rb. }
\end{figure}
The scheme proposed here is depicted as in Fig. 1(a). Two control lasers and a THz wave enter into a cavity with relative phase $\phi_{1}$ due to the closed loop as shown in Fig. 1(b). Those two control lasers drive the atomic transitions $|2\rangle\rightarrow|3\rangle$, $|2\rangle\rightarrow|4\rangle$ with frequency detuning $\Delta_{1}=\omega_{1c}-\omega_{32}$ and $\Delta_{2}=\omega_{2c}-\omega_{42}$. The THz wave as a third control laser couples atomic levels $|3\rangle$ and $|4\rangle$ with a frequency detuning $\Delta_{t}=\omega_{t}-\omega_{43}$. A probe laser ($\omega_{p}$), which has a frequency detuning $\Delta_{c}=\omega_{p}-\omega_{c}$ from cavity mode ($\omega_{c}$), is split by a beam splitter (BS). The two split beams $\alpha_{in,l}$ and $\alpha_{in,r}$ are injected into the opposite sides of the cavity through two beam splitters. With phase control device, a relative phase $\phi_{2}$ exists between $\alpha_{in,l}$ and $\alpha_{in,r}$.  Two detectors are applied to receive the output signal from right and left cavity mirror. The atomic levels $|1\rangle$, $|2\rangle$, $|3\rangle$ and $|4\rangle$ as shown in Fig. 1(b) correspond to $5S_{1/2}\;F=1$, $5S_{1/2}\;F=2$, $5P_{1/2}\;F=1$ and $5P_{3/2}\;F=3$ of $^{87}$Rb, respectively. $\Delta_{ac}=\omega_{c}-\omega_{31}$ is frequency detuning of cavity mode and atomic transition $|1\rangle\rightarrow|3\rangle$. $g\sqrt{N}$ is the collective coupling coefficient of cavity-QED system. $\Omega_{1}(\omega_{1c})$, $\Omega_{2}(\omega_{2c})$ and $\Omega_{t}(\omega_{t})$ are Rabi frequency (angular frequency) of control laser 1, control laser 2 and THz wave. $\omega_{31}$, $\omega_{32}$, $\omega_{42}$ and $\omega_{43}$ are angular frequency of corresponding atomic level spacing. Usually the optical switch is realized based on intensity control of control lasers, here we introduce the phase-control setups to manipulate the intracavity field or output signal.

Under rotating wave approximation, the Hamiltonian based on this scheme is as following,
\begin{equation}
\begin{split}
H\!=\!&\!-\!\hbar\Delta_{c}a^{\dag}a\!-\!\hbar\sum_{j=1}^{N}[(\Delta_{p}\!-\!\Delta_{1})\sigma_{22}^{j}\!+\!(\Delta_{p}\!-\!\Delta_{1}\!+\!\Delta_{2}\!-\!\Delta_{t})\sigma_{33}^{j}\\
&\!+\!(\Delta_{p}\!-\!\Delta_{1}\!+\!\Delta_{2})\sigma_{44}^{j}]\!-\!\hbar\sum_{j=1}^{N}(ga^{\dag}\sigma_{13}^{j}e^{i\phi_{2}}\!+\!\Omega_{1}\sigma_{23}^{j}e^{i\varphi_{1}}\\
&\!+\!\Omega_{2}\sigma_{24}^{j}e^{i\varphi_{2}}\!+\!\Omega_{t}\sigma_{34}^{j}e^{i\varphi_{t}})\!+\!H.C.,
\end{split}
\end{equation}
where $\Delta_{p}=\omega_{p}-\omega_{31}$ is frequency detuning of probe laser and atomic transition $|1\rangle\rightarrow|3\rangle$, $g=\mu_{13}\sqrt{\omega_{c}/2\hbar\varepsilon_{0}V}$ is cavity-QED coupling coefficient, $a^{\dag}$($a$) is the creation (annihilation) operator of cavity photons, $\sigma_{mn}^{j}=|m\rangle\langle n|$ ($m,n=1,2,3,4$) is atomic operator, $\phi_{2}$ is the relative phase of two split probe lasers, $\varphi_{1}$, $\varphi_{2}$ and $\varphi_{t}$ are phases of control laser 1, control laser 2 and THz wave, respectively.

For simplicity, we consider a symmetric Fabry-Perot cavity with field loss rate $\kappa_{l}$ ($\kappa_{r}$) from left (right) cavity mirror, $\kappa_{i}=T_{i}/2\tau$, where $T_{i}$ is the mirror transmission and $\tau$ is the photon round-trip time inside the cavity. Before proceeding further, we need to make some approximations, $\langle{a}\rangle=\alpha$ ($\langle{a^{\dag}}\rangle=\alpha^{*}$) and $\langle a_{in,l}\rangle=\alpha_{in}^{l}$ ($\langle a_{in,r}\rangle=\alpha_{in}^{r}$) are expectation values for the operator of intracavity field and incident light from left (right) mirror, respectively \cite{PRA93/023806}. And we have $\rho_{mm}=\langle\sigma_{mm}\rangle$, $\rho_{12}=\langle\sigma_{12}\rangle e^{i(\phi_{2}-\varphi_{1})}$, $\rho_{13}=\langle\sigma_{13}\rangle e^{i\phi_{2}}$, $\rho_{14}=\langle\sigma_{14}\rangle e^{i(\phi_{2}-\varphi_{1}+\varphi_{2})}$, $\rho_{23}=\langle\sigma_{23}\rangle e^{i\varphi_{1}}$, $\rho_{24}=\langle\sigma_{24}\rangle e^{i\varphi_{2}}$ and $\rho_{34}=\langle\sigma_{34}\rangle e^{i\varphi_{t}}$ for atoms. To explore how two phase-control setups affect the transmission or reflection, we need to solve the following equations of motion for density operator $\rho$ and intracavity light operator $a$.
\begin{equation}
\begin{split}
\dot\rho_{11}\!=\!&\;ig(\alpha^{*}\rho_{13}\!-\!\alpha\rho_{31})\!+\!\frac{\Gamma_{3}}{2}\rho_{33}\!+\!\frac{\Gamma_{4}}{2}\rho_{44},\\
\dot\rho_{12}\!=\!&\;[i(\Delta_{p}\!-\!\Delta_{1})\!-\!\gamma_{12}]\rho_{12}\!-\!ig\alpha\rho_{32}\!+\!i\Omega_{1}\rho_{13}\!+\!i\Omega_{2}\rho_{14},\\
\dot\rho_{13}\!=\!&\;[i(\Delta_{p}\!-\!\Delta_{1}\!+\!\Delta_{2}\!-\!\Delta_{t})\!-\!\gamma_{13}]\rho_{13}\!+\!ig\alpha(\rho_{11}\!-\!\rho_{33})\\
&\!+\!i\Omega_{1}\rho_{12}\!+\!i\Omega_{t}\rho_{14}e^{i\phi_{1}},\\
\dot\rho_{14}\!=\!&\;[i(\Delta_{p}-\Delta_{1}\!+\!\Delta_{2})\!-\!\gamma_{14}]\rho_{14}\!-\!ig\alpha\rho_{34}e^{-i\phi_{1}}\!+\!i\Omega_{2}\rho_{12}\\
&\!+\!i\Omega_{t}\rho_{13}e^{-i\phi_{1}},\\
\dot\rho_{22}\!=\!&\;i\Omega_{1}(\rho_{23}\!-\!\rho_{32})\!+\!i\Omega_{2}(\rho_{24}\!-\!\rho_{42})\!+\!\frac{\Gamma_{3}}{2}\rho_{33}\!+\!\frac{\Gamma_{4}}{2}\rho_{44},\\
\dot\rho_{23}\!=\!&\;[i(\Delta_{2}-\Delta_{t})\!-\!\gamma_{23}]\rho_{23}\!+\!ig\alpha\rho_{21}\!+\!i\Omega_{1}(\rho_{22}\!-\!\rho_{33})\\
&\!-\!i\Omega_{2}\rho_{43}e^{i\phi_{1}}\!+\!i\Omega_{t}\rho_{24}e^{i\phi_{1}},\\
\dot\rho_{24}\!=\!&\;(i\Delta_{2}\!-\!\gamma_{24})\rho_{24}\!-\!i\Omega_{1}\rho_{34}e^{-i\phi_{1}}\!+\!i\Omega_{2}(\rho_{22}\!-\!\rho_{44})\\
&\!+\!i\Omega_{t}\rho_{23}e^{-i\phi_{1}},\\
\dot\rho_{33}\!=\!&\;ig(\alpha\rho_{31}\!-\!\alpha^{*}\rho_{13})\!+\!i\Omega_{1}(\rho_{32}\!-\!\rho_{23})\!+\!i\Omega_{t}(\rho_{34}\!-\!\rho_{43})\\
&\!-\!\Gamma_{3}\rho_{33},\\
\dot\rho_{34}\!=\!&\;(i\Delta_{t}\!-\!\gamma_{34})\rho_{34}\!-\!ig\alpha^{*}\rho_{14}e^{i\phi_{1}}\!-\!i\Omega_{1}\rho_{24}e^{i\phi_{1}}\!+\!i\Omega_{2}\rho_{32}e^{i\phi_{1}}\\
&\!+\!i\Omega_{t}(\rho_{33}\!-\!\rho_{44}),\\
\dot\rho_{44}\!=\!&\;i\Omega_{2}(\rho_{42}\!-\!\rho_{24})\!+\!i\Omega_{t}(\rho_{43}\!-\!\rho_{34})\!-\!\Gamma_{4}\rho_{44},\\
\dot\alpha\!=\!i&\Delta_{c}\alpha\!+\!igN\rho_{13}\!-\!(\kappa_{l}\!+\!\kappa_{r})\alpha\!+\!\sqrt{2\kappa_{l}/\tau}\alpha_{in}^{l}\!+\!\sqrt{2\kappa_{r}/\tau}\alpha_{in}^{r}.
\end{split}
\end{equation}
Here $\phi_{1}=\varphi_{1}-\varphi_{2}+\varphi_{t}$ represents the relative phase of two control lasers and THz wave induced by the closed loop, $\Gamma_{3}=\Gamma_{4}=\Gamma$ is the natural decay rate of excited states $|3\rangle$ and $|4\rangle$, $\gamma_{12}$, much smaller than $\Gamma_{3}$ or $\Gamma_{4}$, is the decoherence rate between ground states $|1\rangle$ and $|2\rangle$, $\gamma_{14}=\gamma_{23}=\gamma_{24}=\Gamma$ stands for the decay rates between corresponding atomic levels and $\gamma_{34}=\sqrt{\Gamma_{3}\Gamma_{4}}=\Gamma$ describes the coupling rate of states $|3\rangle$ and $|4\rangle$ \cite{PRL95/057401}.

By solving above motion equations under steady-state condition, the intracavity field amplitude can be derived as,
\begin{equation}
\alpha=\frac{\sqrt{2\kappa_{l}/\tau}\;\alpha_{in}^{l}+\sqrt{2\kappa_{r}/\tau}\;\alpha_{in}^{r}}{(\kappa_{l}+\kappa_{r})-i\Delta_{c}-i\chi},
\end{equation}
where
\begin{equation*}
\chi\!=\!\frac{g^{2}N(\Omega_{2}^{2}-A*B)}{2\Omega_{1}\Omega_{2}\Omega_{t}\cos{\phi_{1}}-A*\Omega_{t}^{2}-B*\Omega_{1}^{2}-C*\Omega_{2}^{2}+A*B*C}
\end{equation*}
is the susceptibility of atomic medium and $A=\Delta_{p}-\Delta_{1}+i\,\gamma_{12}$, $B=(\Delta_{p}-\Delta_{1}+\Delta_{2})+i\,\Gamma_{4}$, $C=(\Delta_{p}-\Delta_{1}+\Delta_{2}-\Delta_{t})+i\,\Gamma_{3}$. According to input-output relations \cite{G.S.Agarwal2013},
\begin{gather}
\begin{split}
a_{out,l}+a_{in,l}=\sqrt{2\kappa_{l}\tau}\;a,\\
a_{out,r}+a_{in,r}=\sqrt{2\kappa_{r}\tau}\;a,
\end{split}
\end{gather}
the analytical solutions for intracavity light field and output light field through right and left mirror are,
\begin{equation}
\begin{split}
I_{c}&=I_{in}|\frac{\kappa(1+e^{i\phi_{2}})}{\kappa-i\Delta_{c}-i\chi}|^{2},\\
I_{out}^{r}&=I_{in}|\frac{\kappa(1+e^{i\phi_{2}})}{\kappa-i\Delta_{c}-i\chi}-1|^{2},\\
I_{out}^{l}&=I_{in}|\frac{\kappa(1+e^{-i\phi_{2}})}{\kappa-i\Delta_{c}-i\chi}-1|^{2}.
\end{split}
\end{equation}
Here we set $\kappa_{l}=\kappa_{r}=\kappa/2$, $\alpha_{in}^{l}=|\alpha_{in}|e^{i\varphi_{l}}$ and $\alpha_{in}^{r}=|\alpha_{in}|e^{i\varphi_{r}}$, $\phi_{2}=\varphi_{l}-\varphi_{r}$ is the relative phase of two incident probe beams, $I_{in}$ is the input field intensity and $I_{c}$, $I_{out}^{r}$ ($I_{out}^{l}$) are field intensity of intracavity light and output light from right (left) mirror, respectively. In this cavity-QED system, the output light intensity can be manipulated by intensity or frequency of control lasers, but in this paper, we focus on the optical switching based on phase modulation. As indicated in Eqs. (5), by varying the closed-loop phase $\phi_{1}$ existed in $\chi$ we can control medium absorption at different frequency and also we can modulate the output intensity via optical interference induced by the relative phase $\phi_{2}$ between two incident probe lasers $\alpha_{in}^{l}$ and $\alpha_{in}^{r}$. In brief, when medium absorption is decided, we can manipulate the total optical absorption of this system by means of wave interference. In this sense, the output spectra is also an interference fringe of absorption in this scheme.

\section{\normalsize\uppercase{simulated results}}
We consider our system at the threshold of the strong collective-coupling regime ($g^{2}N=\kappa\,\Gamma$) and assume that cavity mode is tuned to the atomic resonance $\Delta_{ac}=0$ (i.e. $\Delta_{c}=\Delta_{p}$). Then we can obtain the output spectra under resonance condition $\Delta_{1}=\Delta_{2}=\Delta_{t}=0$ as in Fig. 2. The parameters used here are $g\sqrt{N}=\Gamma$, $\Omega_{1}=\Omega_{2}=\Omega_{t}=\Gamma$, $\kappa=\Gamma$, $\gamma_{12}=0.001\Gamma$. Fig. 2(a) shows the output field intensity from right mirror $I_{out}^{r}$ varying with changing relative phase $\phi_{2}$, for which the reason is that the reflection from right beam interferes destructively or constructively with the transmission from left beam at the interface of right mirror for different $\phi_{2}$. Due to the frequency-dependence absorption in medium, the output field intensity varies with $\Delta_{p}$ for the same $\phi_{2}$. In Fig. 2(b), the output spectrum is symmetrical about $\phi_{1}=\pi$, which is based on the similar principle with Ref. \cite{CPB26/074207}. And it is easy to be predicted from the susceptibility of atomic medium. Here the output intensity is controlled by operating either the setup of phase control $\phi_{1}$ or the phase control $\phi_{2}$. Different from traditional methods, we would like to explore what the situation will be if both $\phi_{1}$ and $\phi_{2}$ are applied to manipulate output light field. With the phase-control setup of $\phi_{1}$, the medium absorption can be modulated. Via phase-control setup of $\phi_{2}$, the photon escape from cavity can be manipulated through the optical interference at the interface of cavity mirror. Therefore, combining the controlling of medium properties with optical interference, we will focus on total phase control of output spectra in the following.
\begin{figure}[!hbt]
\centering
\includegraphics[width=6cm]{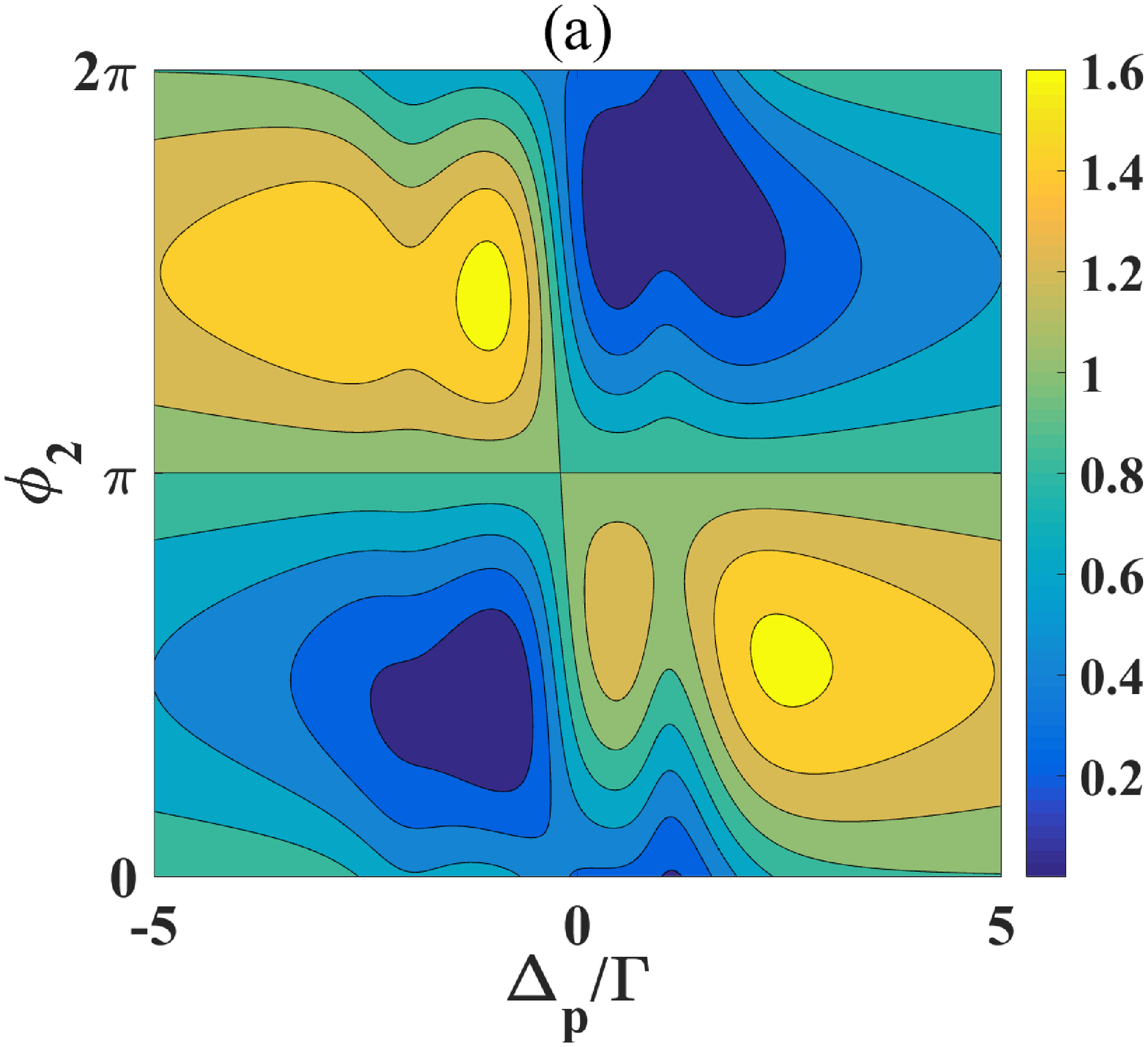}
\includegraphics[width=6cm]{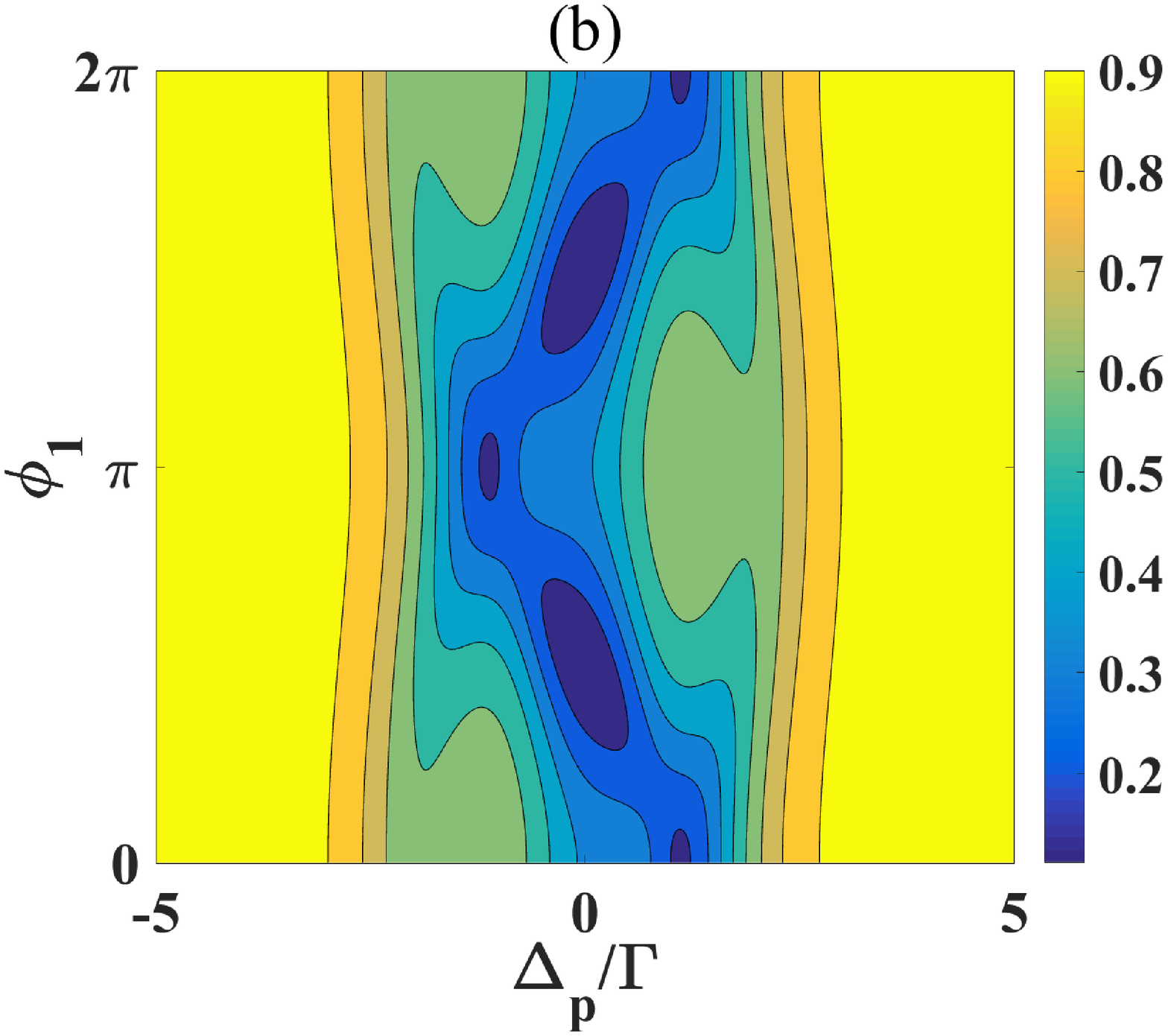}
\caption{Contour plots of $I_{out}^{r}$ as function of frequency detuning $\Delta_{p}$ and (a) relative phase $\phi_{2}$, (b) closed-loop phase $\phi_{1}$. The parameters are (a) $\phi_{1}=0$ and (b) $\phi_{2}=0$.}
\end{figure}
\begin{figure*}[htbp]
\centering
\includegraphics[width=5.5cm]{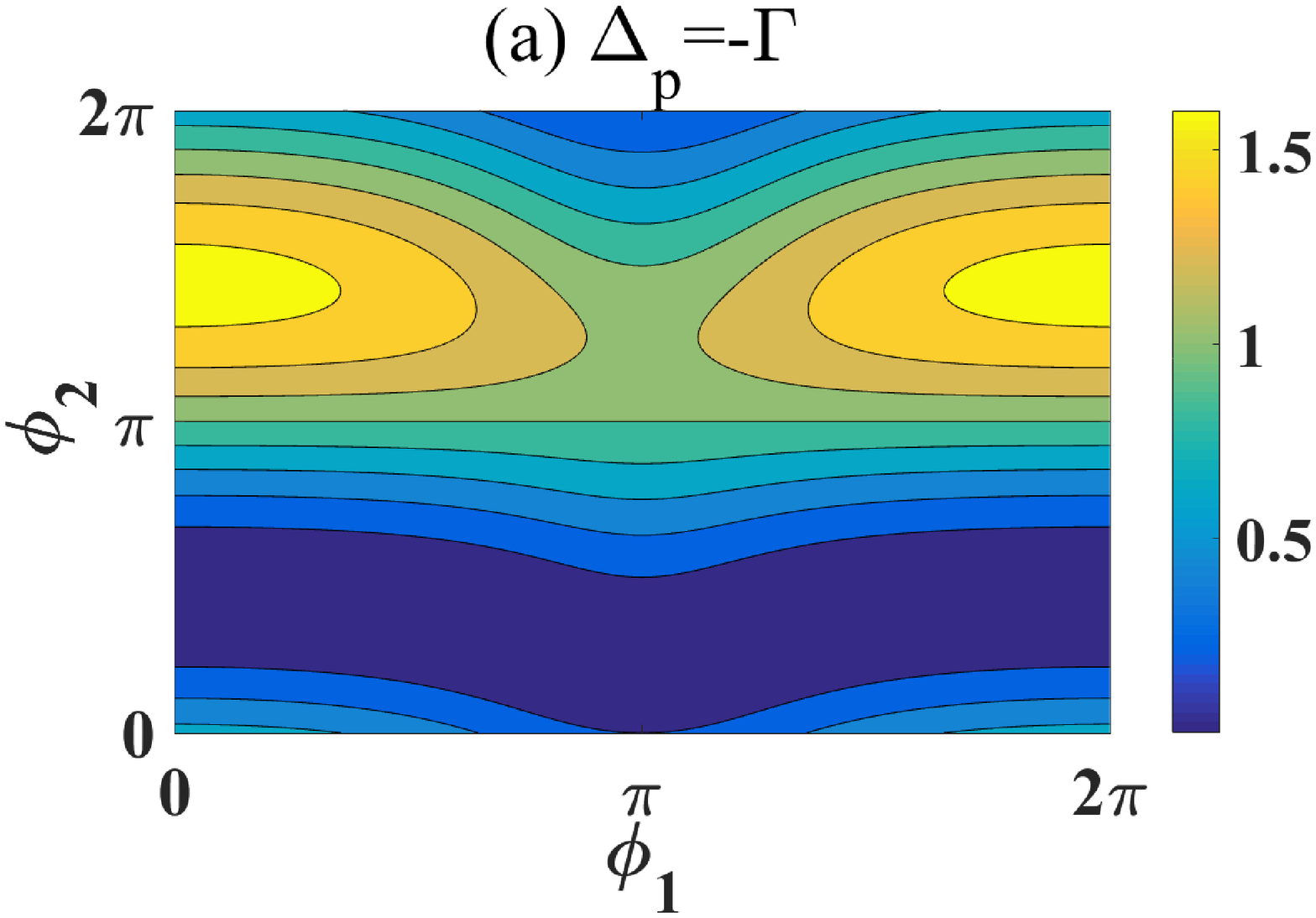}
\includegraphics[width=5.5cm]{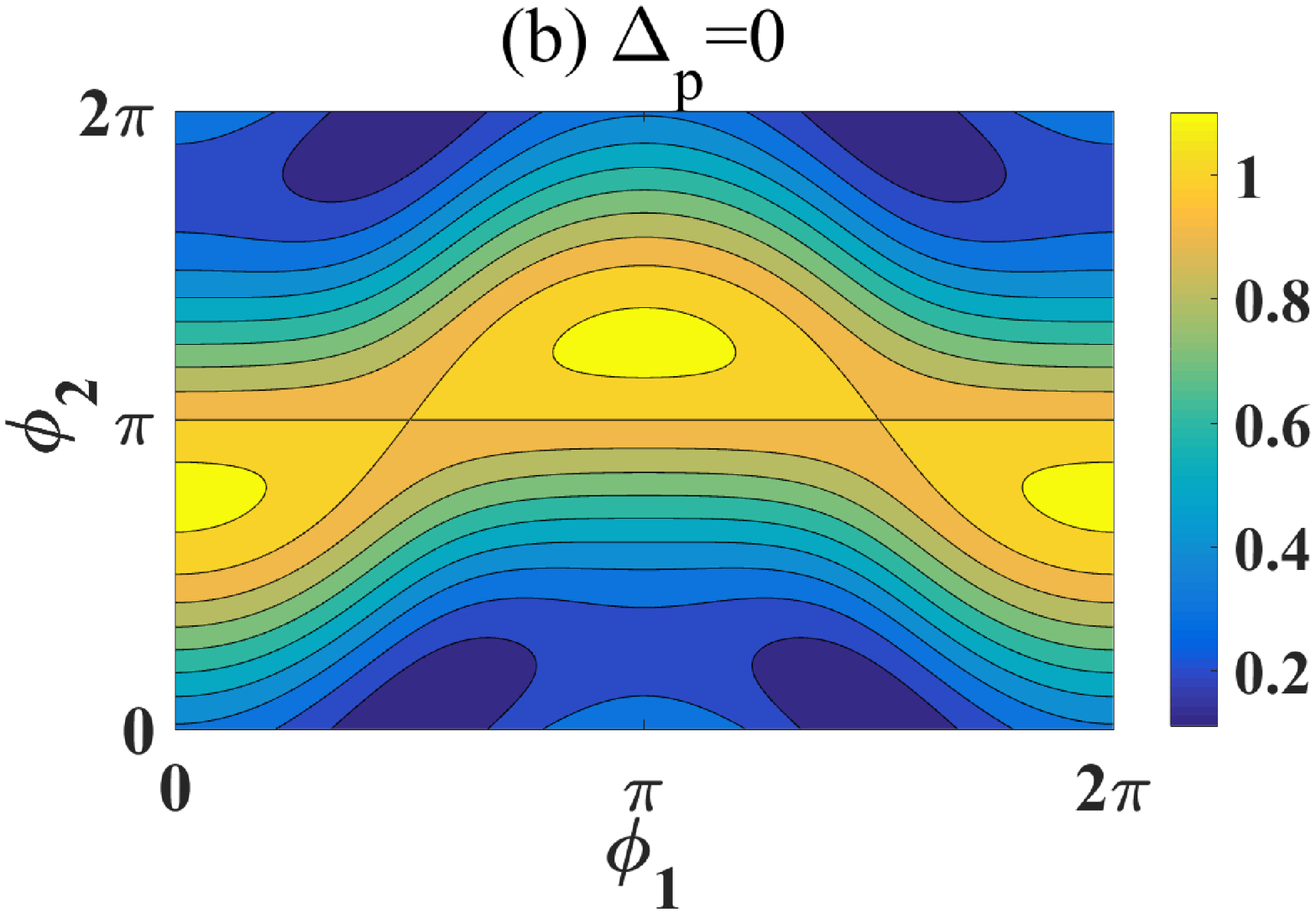}
\includegraphics[width=5.5cm]{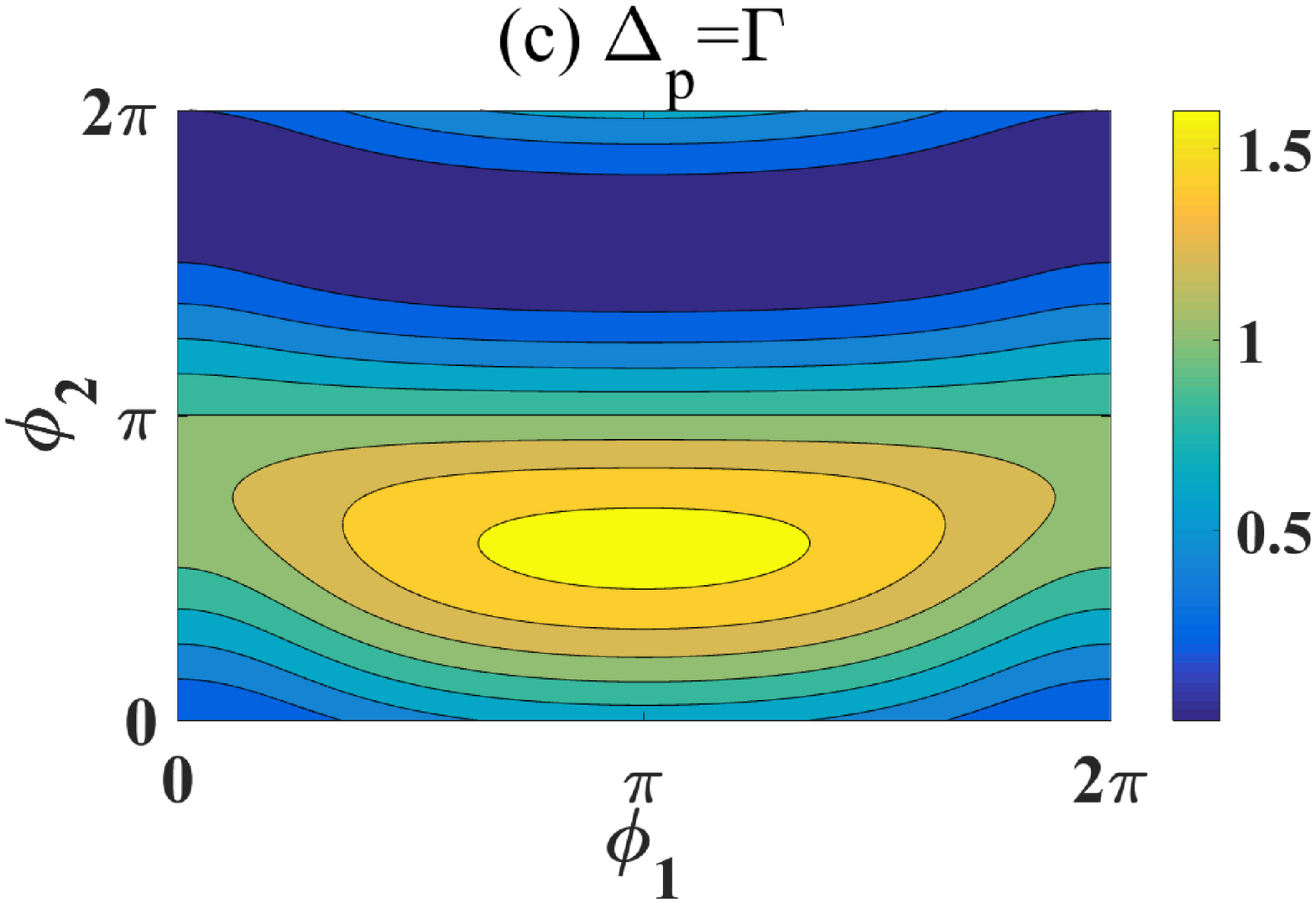}
\caption{Contour plots of $I_{out}^{r}$ as function of closed-loop phase $\phi_{1}$ and relative phase $\phi_{2}$ for three representative frequency detuning (a) $\Delta_{p}=-\Gamma$, (b) $\Delta_{p}=0$ and (c) $\Delta_{p}=\Gamma$. The parameters here are $g\sqrt{N}=\Omega_{1}=\Omega_{2}=\Omega_{t}=\Gamma$, $\gamma_{12}=0.001\Gamma$.}
\end{figure*}

Here we choose three representative frequency $\Delta_{p}=-\Gamma$, $\Delta_{p}=0$ and $\Delta_{p}=\Gamma$ to analyze the phase control of output field intensity. We show the contour plots of output field intensity$I_{out}^{r}$ versus $\phi_{1}$ and $\phi_{2}$ in Fig. 3. They present the symmetry about $\phi_{1}$ same with that shown in Fig. 2(b) for all three frequency detuning. While the situation for $\phi_{2}\rightarrow 0\sim\pi$ is very different from that for $\phi_{2}\rightarrow \pi\sim 2\pi$, it shows the optical interference relevant to frequency mode. When it is at negative probe-frequency detuning as in Fig. 3(a), the photon-escape enhancement appears at $\phi_{2}\rightarrow\pi\sim2\pi$. Different absorption causes the output intensity reducing with changing $\phi_{1}$ from $0$ to $\pi$ or from $2\pi$ to $\pi$. When the probe frequency is resonant with atomic transition $|1\rangle\rightarrow|3\rangle$ (Fig. 3(b)), the enhancement of output intensity disappears and the max output intensity appears when $\phi_{2}$ is near $\pi$. That is because the closer to $\pi$ for $\phi_{2}$, the closer to 0 for intacavity field intensity as indicated in Eq. (5), namely $\phi_{2}=\pi$ is the condition for perfect transmitter/reflector. According to Fig. 2(b), it is easy to understand why the max output intensity for $\Delta_{p}=0$ is at $\phi_{1}\approx0$, $\pi$ and $2\pi$ respectively. While for positive probe-frequency detuning as in Fig. 3(c), the output enhancement is generated as changing $\phi_{2}$ from 0 to $\pi$ and the enhancement is weakening with changing $\phi_{1}$ from $\pi$ to 0 or $2\pi$ which is opposite to Fig. 3(a). For $\phi_{2}$ varying from 0 to $2\pi$, Fig. 3 presents remarkable sign of optical interference and the interference fringes become more similar with general fringes when the probe frequency is under non-resonant case. The nonuniform intensity distribution of interference fringes is due to the nonlinear absorption and multiple diffraction of input light and transmission (reflection) \cite{Science331/889,PRL55/2696}.

\begin{figure*}[htbp]
\centering
\includegraphics[width=4.5cm]{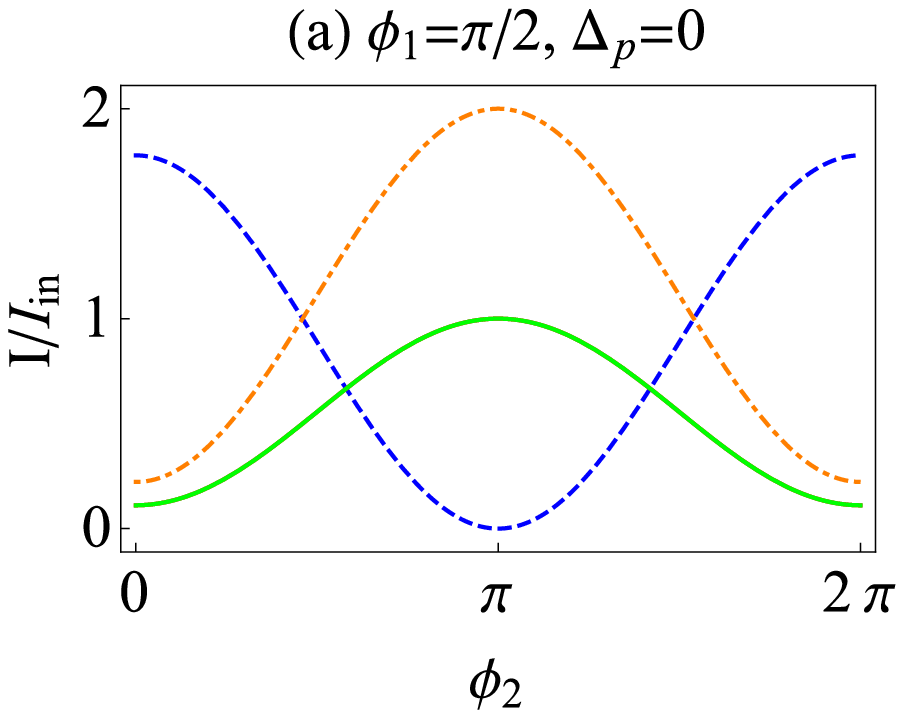}
\includegraphics[width=4.5cm]{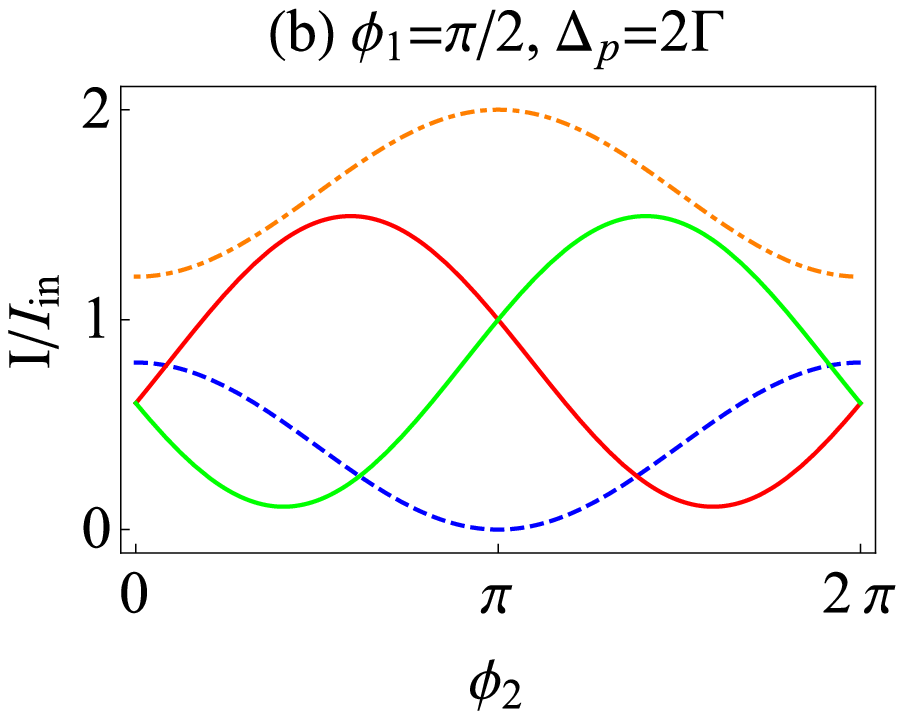}
\includegraphics[width=4.5cm]{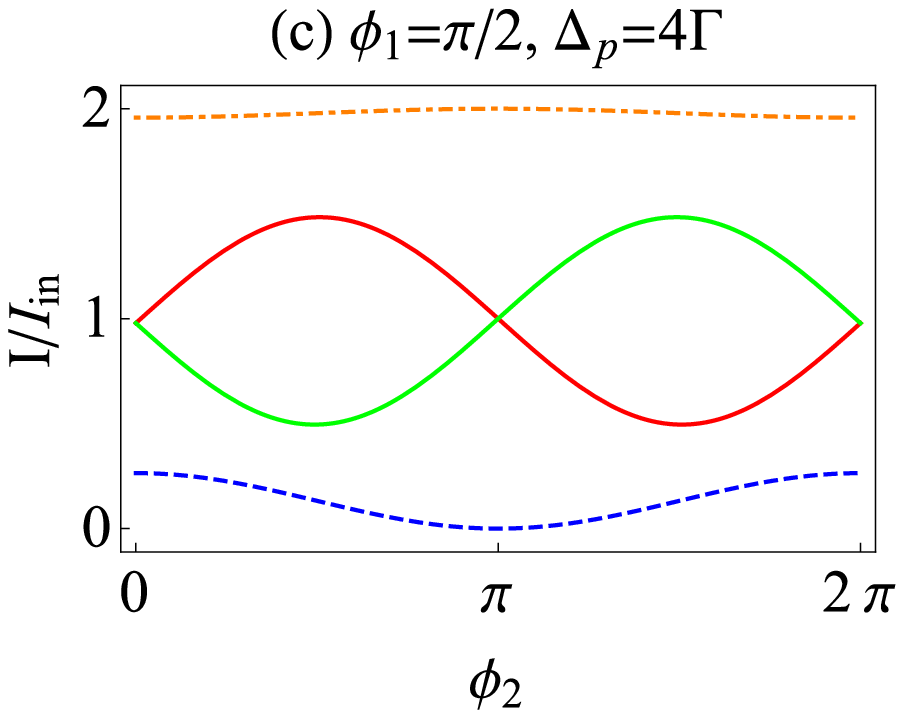}
\includegraphics[width=3.5cm]{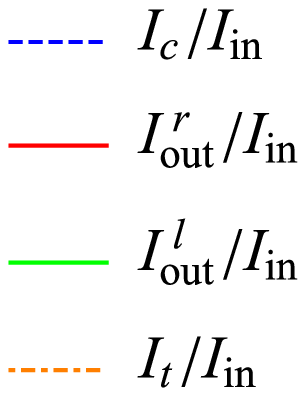}
\caption{Field intensity ratio versus incident relative phase $\phi_{2}$ for intracavity field ($I_{c}/I_{in}$), output field from right or left mirror ($I_{out}^{r}/I_{in}$ or $I_{out}^{l}/I_{in}$) and total output field ($I_{t}/I_{in}$) . The parameters are $\phi_{1}=\pi/2$ and (a) $\Delta_{p}=0$, (b) $\Delta_{p}=2\Gamma$, (c) $\Delta_{p}=4\Gamma$. The other parameters are the same as in Fig. 3. }
\end{figure*}
Since the output spectra are always symmetrical about $\phi_{1}=\pi$ with modifying relative phase $\phi_{2}$ no matter what the probe frequency is, we set closed-loop phase setup to make $\phi_{1}=0, \;\pi/2, \;\pi$ in turn, then we analyze the intracavity light field, the total output intensity, the output light field from right and left mirror for frequency resonance and detuning far from resonance. Here we present the result for $\phi_{1}=\pi/2$ in consideration that the variation tendency for $\phi_{1}=0$ and $\phi_{1}=\pi$ are similar with that for $\phi_{1}=\pi/2$ except one phenomenon that $I_{out}^{r}=I_{out}^{l}$ for whole range of $\phi_{2}$ ($0\rightarrow2\pi$) at $\Delta_{p}=0$ as shown in Fig. 4(a). The situation for negative frequency detuning is not presented here, because the only difference is that the output field from right (left) mirror for negative frequency detuning has fixed phase delay with the situation at positive frequency detuning.

In Fig. 4, when $\phi_{2}=\pi$, in which the perfect transmitter/reflector condition is satisfied, intracavity light intensity is always zero, the output field intensity from right mirror is the same with that from left and total output intensity ratio is always in the maximum value 2, which manifest the frequency independence as well as medium irrelevance of field intensity for $\phi_{2}=\pi$. In Fig. 4(a), when $\phi_{2}=0\;(2\pi)$ the incident probe photon is nearly all trapped in the cavity. The cavity is with fixed lossy rate from two sides and the perfect photon trapping condition \cite{PRA92/023824} is not met in this scheme, as a result, there are tiny output light from two cavity mirrors at probe frequency resonance. When increasing the frequency detuning $\Delta_{p}$, the total output intensity becomes lower sensitive to the variation of $\phi_{2}$ and the phase delay between $\I_{out}^{r}$ and $I_{out}^{l}$ is visible. When $\Delta_{p}$ is up to $4\Gamma$, the total output intensity is almost insensitive to the modulation of $\phi_{2}$ and the output intensity from two cavity mirrors exist a $\pi$ delay dependent on $\phi_{2}$. Fig. 4 shows the property of wave interference relevant to medium absorption in this scheme. That's why our scheme may be acted as an absorption interferometer. It can also be used to realize state transfer between two output channels via phase modulation of $\phi_{2}$. Moreover the manipulation of field intensity based on frequency detuning makes this scheme a subtle frequency filter via phase control.
\begin{figure*}[!hbt]
\centering
\includegraphics[width=4.8cm]{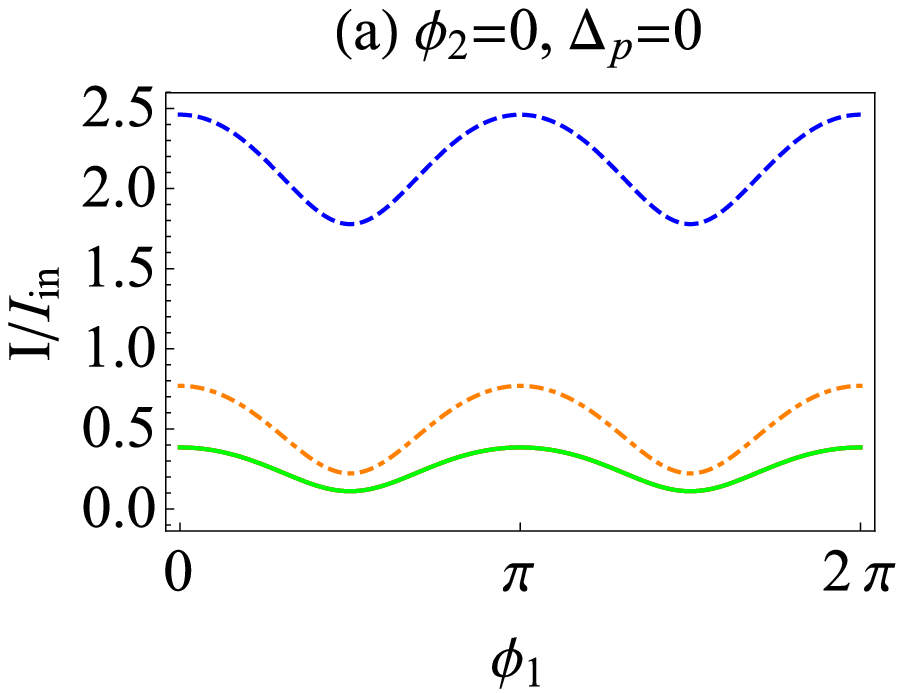}
\includegraphics[width=4.8cm]{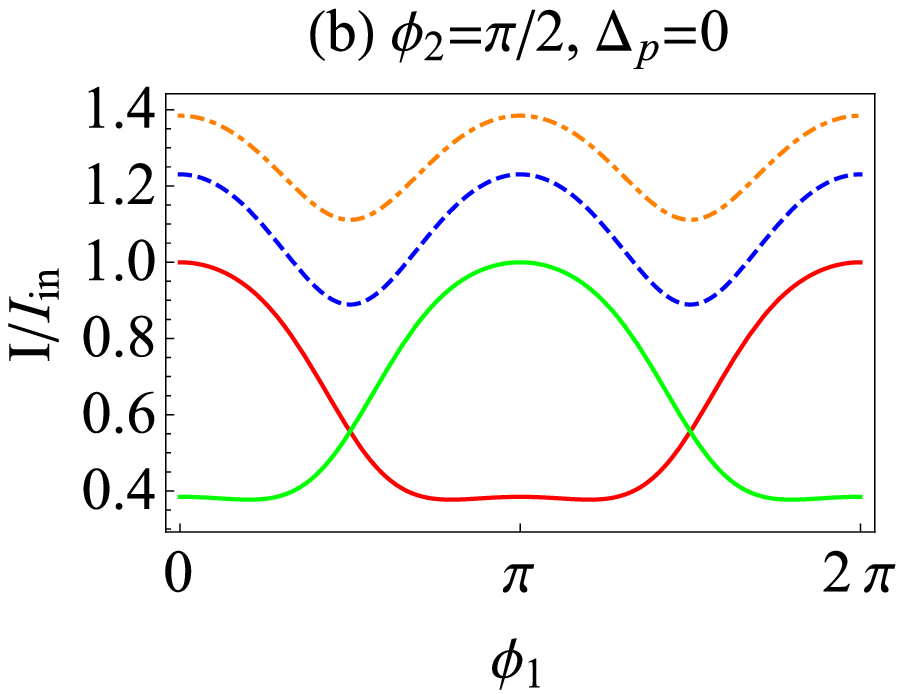}
\includegraphics[width=4.8cm]{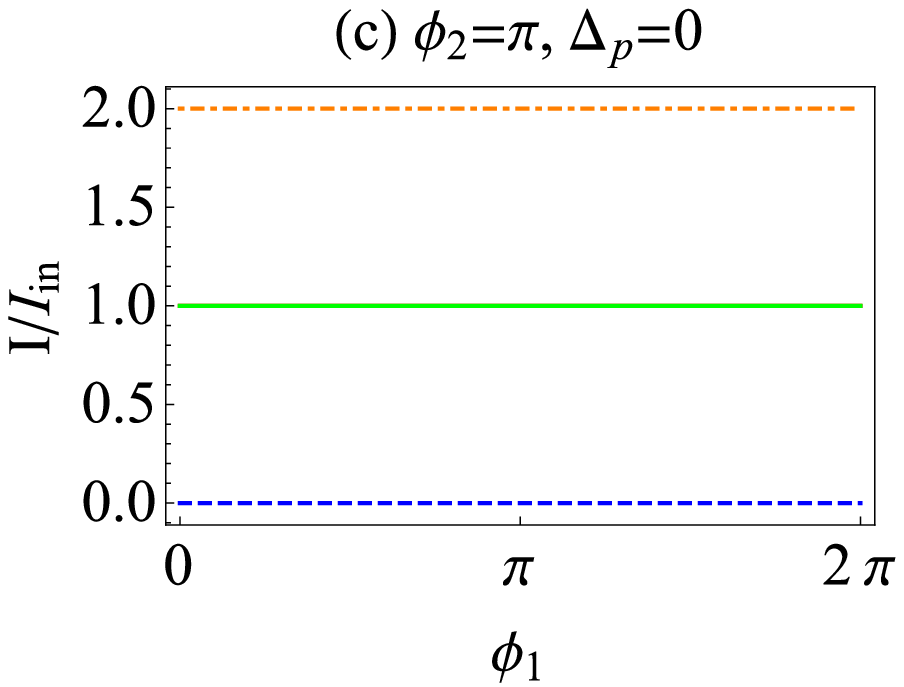}
\includegraphics[width=3.5cm]{fig4l}
\caption{With probe frequency tuned with atomic transition $|1\rangle\rightarrow|3\rangle$ ($\Delta_{p}=0$), the field intensity ratio versus closed-loop phase $\phi_{1}$ for (a) $\phi_{2}=0$, (b) $\phi_{2}=\pi/2$ and (c) $\phi_{2}=\pi$. The other parameters here are the same as in Fig. 3.}
\end{figure*}

If probe laser is tuned at resonance ($\Delta_{p}=0$), the output intensity variation with closed-loop phase $\phi_{1}$ is very distinct among $\phi_{2}=0$, $\pi/2$ and $\pi$ as in Fig. 5. The output intensity $I_{out}^{r}$ and $I_{out}^{l}$ can not be distinguished from each other with changing closed-loop phase $\phi_{1}$ when relative phase $\phi_{2}$ is even order of $\pi/2$ (e.g. $\phi_{2}=0$ in Fig. 5(a) and $\phi_{2}=\pi$ in Fig. 5(c)). This phenomenon also happens when probe laser is not at resonance. Only when $\phi_{2}$ is non-even order of $\pi/2$, the complex amplitude will carry different phase information for output field through right and left mirror,  and the intensity of output light through right mirror is separated from that through left mirror by a $\pi$ phase delay. Here we show one special value $\phi_{2}=\pi/2$ in Fig. 5(b). By increasing $\phi_{2}$ from 0 to $\pi$, Fig. 5 shows that the photon escape from cavity is enhanced at probe frequency resonance. The light intensity is periodic modulated by $\phi_{1}$ before $\phi_{2}=\pi$. When $\phi_{2}=0$ (Fig. 5(a)), the output intensity is at minimum value due to destructive interference, and it can be modulated by $\phi_{1}$ based on this special atomic system where the medium absorption can be manipulated by the closed-loop phase.  When $\phi_{2}=\pi$ (Fig. 5(c)), optical interference plays the main role and the completely constructive interference in this mode at the interface of cavity mirrors makes the output intensity maximum and the intracavity field equal to zero. It shows a perfect transmitter/reflector with closed-loop-phase insensitivity. Therefore our scheme can be applied to fulfill a switching between coherent photon absorber (CPA) and a perfect photon transmitter or reflector based on phase modulation.

\section{\normalsize\uppercase{conclusions}}
In conclusion, we have analyzed the optical field intensity at the threshold of strong collective-coupling regime in the four-level atom-cavity system here. Due to the closed loop, there will be a relative phase $\phi_{1}$ between three control lasers in susceptibility of atomic medium, thus we can manipulate the intracavity and output field intensity by controlling the medium absorption via $\phi_{1}$. Since we use two coherent input beams with relative phase $\phi_{2}$, the analytical solutions for both intracavity and output field will contain two phase factors $\phi_{1}$ and $\phi_{2}$. Therefore, besides the absorption controlling by closed-loop phase $\phi_{1}$, the optical interference dependent on relative phase $\phi_{2}$ can be applied to manipulate the field intensity.

Via the contour plots versus $\phi_{1}$ and $\phi_{2}$, we analyze the interaction of medium absorption and wave interference, and we make a propose about absorption interferometer based on phase modulation. We derive the perfect transmitter/reflector condition, then we realize the operation of switching between perfect transmitter/reflector and non-perfect CPA theoretically. And the phase delay about $\phi_{2}$ between $I_{out}^{r}$ and $I_{out}^{l}$ provides us a way to fulfill frequency filtering. By controlling closed-loop phase $\phi_{1}$, the intracavity field can be modulated periodically. Therefore the scheme proposed here can be applied to operate light field more delicately in cavity-QED system.

\section*{\normalsize\uppercase{acknowledgment}}
We acknowledge support from National Natural Science Foundation of China under Grant No. 11174109.

\end{document}